\documentclass[prd,aps,preprint,epsfig,floats]{revtex4}
\usepackage{graphics}
\usepackage{epsfig}

\usepackage{graphicx}

\begin{document}

\title{\Large\bf Complex CKM from Spontaneous CP Violation Without
 Flavor Changing Neutral Current}
\author{\bf G. C. Branco$^{1,2}$ and R.N. Mohapatra$^{1,3,4}$ }

\affiliation{$^1$ Theoretische Physik, Tech. Univ. Muenchen,
Garching, Germany\\
$^2$ Departmento de Fisica and CFTP, Instituto Superior T\'ecnico, Lisboa, 
Portugal\\
 $^3$Department of Physics, University of Maryland, College
Park, MD 20742, USA\\
$^4$ Sektion Physik, Universitaet Muenchen, Muenchen, Germany }

\date{July, 2006}

\begin{abstract}
We analyse the general constraints on unified gauge models with spontaneous CP 
breaking that satisfy the conditions that (i) CP violation in the 
quark sector is described by a realistic complex CKM matrix, and 
(ii) there is no significant flavor changing neutral current effects in 
the quark sector. We show that the crucial requirement in order to 
conform to the above conditions is that spontaneous CP breaking occurs 
at a very high scale by complex vevs of standard model singlet Higgs 
fields. Two classes of models are found, one consisting of pure Higgs 
extensions and the other one involving fermionic extensions of the 
standard model. We give examples of each class and discuss 
their possible embeddings into higher unified theories. One of the models 
has the interesting property that spontaneous CP violation is triggered by 
spontaneous P violation, thereby linking the scale of CP violation to 
the seesaw scale for neutrino masses.
 \end{abstract}
\maketitle
\section{Introduction}
It is now becoming increasingly clear that the dominant contribution to
low energy CP violation arises from the complex CKM matrix which 
parameterizes the weak quark current coupling to the W-boson. Indeed 
the recent measurement \cite{gamma} of the angle 
$\gamma=-Arg(V_{ud}V_{cb}V^*_{cd}V^*_{ub})$ provides evidence\cite{botella}
for a complex CKM matrix even if one allows for New Physics (NP) 
contributions to $B_d-\bar{B}_d$ mixing and $B_s-\bar{B}_s$ mixings.

However, this cannot be the full story of CP violation in elementary particle
interaction\cite{book} since it is believed that the explanation of the 
only cosmic manifestation of CP nonconservation i.e. the asymmetry
between matter and anti-matter must come from sources other than
the CKM CP violation; similarly the solution to the QCD $\theta$
problem may also imply new forms of CP violating interactions. Moreover, 
there is the fundamental question of the origin and nature of CP 
violation and its relation to other constituents and forces.

Even before the full story of CP violation is clear, one can ask
the question as to whether the observed CKM CP violation is
spontaneous in origin\cite{lee} or intrinsic to the Yukawa
couplings in the theory. This question has nontrivial cosmological
implications since spontaneous CP violation will lead to domain
walls and in order to avoid conflict with observations such as
WMAP data, one must have the scale of this breaking to be above that
of the inflation reheating, thus imposing constraints on both
cosmological as well as particle physics aspects of models.

In practical construction of models with spontaneous CP breaking,
one must have one or more Higgs fields to have complex
vevs\cite{lee}. It is obvious that implementing this requires
extending the standard model, by having either more Higgs/or
fermion fields plus Higgs because gauge invariance allows no room
for Higgs vevs to be complex in the standard model. Furthermore,
since spontaneous CP violation (SCPV) requires nontrivial
constraints on the realistic gauge models, it is not surprising
that the process of implementing it can lead to unpleasant side
effects. One such unpleasant effect is the plethora of flavor
changing neutral current (FCNC) effects induced in the process of
obtaining spontaneous CP breaking.
 
Therefore, the challenge in constructing realistic models with 
spontaneous CP violation is twofold:

i) One should achieve genuine spontaneous CP violation and assure that the 
vacuum phase does lead to a non-trivially complex CKM matrix. This is not 
an easy task since CP invariance of the Lagrangian requires the Yukawa 
couplings to be real.

ii) One should find a natural suppression mechanism for FCNC in the Higgs 
sector. Again, this is a challenging task, since there is in general a 
close connection \cite{gcb} between the appearance of FCNC and the 
possibility of generating a complex CKM matrix through CP violating 
vacuum phases.

The above link between SCPV and FCNC can be seen by considering
a two Higgs extension $(\phi_{1,2})$ of the standard model to
implement SCPV. It is well known (and we repeat the derivation in
sec.II and in Appendix A) that general two Higgs models have FCNC mediated 
by neutral Higgs fields.
In order to suppress these FCNC effects one may consider two 
possibilities. One consists of the introduction of extra symmetries which 
eliminate FCNC and guarantee natural flavour conservation ( NFC ) 
\cite{glashow} in 
the Higgs sector. It is well known that the introduction of such 
symmetries in the two Higgs doublet framework eliminates the possibility 
of having spontaneous CP violation \cite{gcb} . With three Higgs doublets 
one can 
have NFC and yet achieve spontaneous CP violation but the resulting CKM
matrix is real, in contradiction with recent data. Above we have 
considered the case where FCNC are avoided through the introduction of
extra symmetries, not by fine-tuning. It has been shown that even if one 
considers elimination of FCNC through fine-tuning, for three generations
one cannot generate a realistic complex CKM matrix \cite{grimus}.
The other possibility for suppressing FCNC effects is by choosing a large 
mass for the neutral Higgs which violate flavour. Indeed the strength of
FCNC effects  is proportional to $1/M^2_{H}$ where $H$ denotes the new
neutral Higgs field (we will denote the standard model Higgs by
$h$). So clearly, suppression of FCNC effects require that $M_H$
become very large . On the other hand, as we show below, the magnitude of 
the CP phase (denoted by $\delta$ in the text) in this model is given by 
$\delta \sim\frac{M_{W}}{M_H}$ so that as $M_H\to $ very large, $\delta\to 0$
and the theory becomes almost CP conserving. Note that to obtain
CKM CP violation, we need $\delta \sim 1$. We will thus show that in the 
context of models with SCPV at the electroweak scale, it is not possible 
to obtain a complex CKM matrix while suppressing FCNC effects. 
In this class of SCPV models, obtaining a large CP phase and having 
significant FCNC seem to go together.

In this paper, we discuss the conditions under which this
connection can be avoided. We point out that the crucial point is to
have CP broken at a high energy scale. We present two classes of models: 
one where the extension involves only the Higgs sector of the standard
model and another one which involves the fermion sector as well. In
the latter case, there is a small departure from unitarity of the
CKM matrix.

Several of the models we discuss have already been considered in the
literature. We present a systematic classification of these  models,
adding some new ones and sharpening the connection between SCPV and
FCNC. In particular, we present criteria for constructing realistic SCPV
models free of FCNC constraints.

This paper is organized as follows: in sec. II, we discuss the connection 
between SCPV and FCNC in  doublet Higgs extension of the SM. In sec. III, 
we discuss spontaneous CP breaking at high scale in a pure Higgs 
extension and show how one can avoid the FCNC effects in this case. In 
sec. IV, we present a fermionic extension of the SM with spontaneous CP 
breaking at high scale. In sec. V, we discuss these two classes models 
into a left-right model and discuss two models one of which has the 
interesting property that spontaneous CP violation is triggered by 
spontaneous P violation. In sec. VI, we briefly comment on
how our ideas can be extended to supersymmetric models and finally in sec. 
VII, we present our conclusions. In the appendices A and B we present a 
detailed demonstration of the results of sec. in sec. II and III.

\section{Two Higgs Doublet Model for SCPV and FCNC}
The simplest extension of the standard model that can accommodate
spontaneous CP violation is the two Higgs doublet model. If we
denote the two Higgs doublets as $\phi_{1,2}$, and define
$V_0(x,y)=-\mu^2_1 x-\mu^2_2 y+\lambda_1 x^2 + \lambda_2 y^2+\lambda_3 xy$, 
we can write the potential as follows:

\begin{eqnarray}
V(\phi_{1,2})~=~V_0(\phi^\dag_1\phi_1,
\phi^\dagger_2\phi_2)+V_{12} \label{eq1}
\end{eqnarray}
where
\begin{eqnarray}
V_{12}(\phi_1, \phi_2)~=~\mu^2_{12}\phi^\dag_1\phi_2
 +~\lambda_4(\phi^\dag_1\phi_2)^2 +\lambda_5
\phi^\dag_1\phi_2 \phi^\dag_1\phi_1 +\lambda_6
\phi^\dag_1\phi_2 \phi^\dag_2\phi_2 ~+~h.c.\\ \nonumber
+~\lambda'_3\phi^\dag_1\phi_2 \phi^\dag_2\phi_1~
 \label{eq2}
\end{eqnarray}
We can now write down the potential in terms of the electrically
neutral components of the doublets. It looks exactly the same as the
above potential as long as we understand the various fields as the
neutral components of the fields.

In order to discuss spontaneous CP violation\cite{haber}, we look for a
minimum of the form:
\begin{eqnarray}
<\phi_1>~=~\pmatrix{0\cr \frac{1}{\sqrt{2}}v_1};
<\phi_2>~=~\pmatrix{0\cr \frac{1}{\sqrt{2}}v_2e^{i\delta}};
\end{eqnarray}
The potential at this minimum looks like
\begin{eqnarray}
V(v^2_1,v^2_2,\delta)~=~V_0(v^2_1,v^1_2)~+~\frac{1}{4}\lambda'_3 v^2_1v^2_2\\
\nonumber +\mu^2_{12}v_1v_2
cos\delta+\frac{1}{2}\lambda_4v^2_1v^2_2cos2\delta~+~\frac{1}{2} 
(\lambda_5v^2_1+\lambda_6
v^2_2)v_1v_2 cos\delta
\end{eqnarray}
The three extremum equations are:
\begin{eqnarray}
\left[-\mu^2_1+\lambda_1v^2_1+(\lambda_3+\lambda'_3)v^2_2+ 
\lambda_4v^2_2sin 2\delta \right]v_1+
v_2\left[\mu^2_{12} cos\delta+\frac{1}{2}(3\lambda_5
v^2_1+\lambda_6v^2_2)cos\delta\right] =0\\
 \left[-\mu^2_1+\lambda_2v^2_2+\frac{1}{2}(\lambda_3+\lambda'_3)v^2_1+ 
\lambda_4v^2_1sin 2\delta \right]v_2+
v_1\left[\mu^2_{12} cos\delta+\frac{1}{2}(\lambda_5
v^2_1+3\lambda_6v^2_2)cos\delta\right] =0\\
-sin\delta \left[\mu^2_{12}v_1v_2+2\lambda_4
v^2_1v^2_2cos\delta+v_1v_2 (\lambda_5 v^2_1+\lambda_6 v^2_2)\right]~=~0
\end{eqnarray}
Now let us study the implications of the extremum equations for
SCPV and FCNC. Writing the Yukawa couplings as ${\cal
L}_Y~=~\sum_{a,b;i} h^{u,i}_{ab}(\bar{Q}_{La}\phi_{i}u_{R,b}+
u\rightarrow d) + h.c.$, it is straightforward to see that in general
there will be FCNC mediated by neutral Higgs. We will consider next 
two possibilities for suppressing these FCNC. One involves the 
introduction of extra symmetries in order to implement Natural Flavour
Conservation (NFC) \cite{glashow} in the Higgs sector; the other considers 
the possibility of making very heavy the neutral Higgs which mediate FCNC. 
We will see that both possibilities do not work as far as generating a 
viable complex CKM matrix, but the discussion is 
useful in order to motivate the breaking of CP at a high energy scale 
which will be considered  in sections 3 and 4.

\subsection {Eliminating FCNC through extra symmetries}

It is well known that it is possible to avoid FCNC by introducing for 
example a $Z_2$ symmetry which restricts the Yukawa couplings so that 
only one  
Higgs doublet gives mass terms to the down quarks while the other doublet 
gives mass to the up quarks. However, it has been shown \cite{gcb} that 
the same 
symmetry which leads to these selective Yukawa couplings prevents the 
occurrence spontaneous CP breaking. A possible way out of this difficulty 
involves the introduction of a third Higgs doublet. In this case it is 
possible to obtain a CP violating vacuum \cite{gcb}
but the 
CKM matrix is real, in conflict with the recent experimental findings. 
The reason why CKM matrix is real in this case has to do with the fact 
that due to the selective Yukawa couplings, the vacuum phase which 
appears in the quark mass matrices can be eliminated by rephasing right 
handed quark fields.

\subsection { Suppressing FCNC effects through large Higgs masses } 

It is straightforward to see that we could
diagonalize one set of Yukawa couplings $h^{u,d,1}$ so that the
neutral Higgs ($h$) coming from the doublet $\phi_1$ has flavor
conserving couplings whereas that from $\phi_2$ ($H$) has flavor
violating couplings. In general of course the two neutral Higgs
fields mix and therefore the $h^{u,2}$ coupling which in the
symmetry limit involves only the $H$ Higgs field will have an
admixture of the light Higgs $h$ but mixing is is always
proportional to the mass ratio $m^2_h/M^2_H$ assuming $M_H\gg
m_h$.

Thus FCNC processes will arise via the tree level exchange of $H$
boson and will be proportional to $M^{-2}_H$ and a contribution
from the mixing term which due to the mixing will also have the
same kind of power dependence on $M_H$. Therefore in order to
suppress FCNC interactions, we must demand that $M_H$ be very
large. This can be achieved by making $-\mu^2_2
> 0$ and $|\mu^2_2|\gg v_{wk}$.
Let us now study  Eq. (6): this equation tells us
 the scale of the vev $v_2$  which depends on the scale of the mixing 
term $\mu_{12}$.
(Note that getting the correct weak scale fixes $\mu^2_1$ to be of
order $v_{wk}$ and stopping FCNC tells us that $|\mu^2_2|\gg
v^2_{wk}$ but so far $\mu^2_{12}$ remains a free parameter.) We
have two cases: (i) $\mu^2_{12}\sim v^2_{wk}$ and (ii) $\mu^2_{12}\sim
M^2_H \sim |\mu^2_2|\gg v^2_{wk}$. In case (i), it is easy to see
using the middle equation above that:
\begin{eqnarray}
v_2\sim \lambda_5\frac{v^3_1}{|\mu^2_2|}\ll v_1
\end{eqnarray}
i.e. the vev of $<\phi_2$ is highly suppressed in the limit of no
FCNC. Note that the mass of the second neutral Higgs is not of
order $v_2$ since in this case the vev is induced by a tadpole
like diagram. Substituting this small value of $v_2$ in Eq.(c), we
then see that for natural values of the parameters ($\lambda_i$),
the only solution for the CP violating phase is $\delta=0,\pi,
...$ .

On the other hand in case (ii), $v_2\sim v_{wk}$ but
equation (7) above tells us that in this case also the expression in
the bracket cannot give a nonzero $\delta$ since
$\mu^2_{12}v_1v_2\gg 2\lambda_4 v^2_1v^2_2$ and the term within
the bracket cannot vanish meaning that sin$\delta=0$ and hence no
SCPV.

 We therefore conclude that in this simple model, the requirement of
suppression of the neutral current effects implies no SCPV. The
main point is that to get a large enough SCPV phase, Eq(7) tells
us that $v_2$ must be comparable in magnitude to $v_1$. For this
to happen, we must have $|\mu^2_2|\sim v^2_{wk}$  which again
means that there must be large FCNC effects at low energies.

The above result can also be seen as follows: In a two Higgs
doublet theory, one can change the basis of Higgs bosons to pass
to a basis where the new doublets are
$\Phi_1~=~(v_2e^{i\delta}\phi_1-v_1\phi_2)/\sqrt{v^2_1+v^2_2}$ and
$\Phi_2$ is the orthogonal combination to $\Phi_1$, where we have
anticipated the vevs of the fields in the original basis, as
discussed above. Now we see that $<\Phi_1>=0$ while $<\Phi_2>\neq
0$ and it leads to the same mass matrices for quarks as before.
Now we can choose parameters of the Higgs potential such that the
mass of $\Phi_1$ is very large to avoid FCNC effects. In this
case, the effective theory below the mass of $\Phi_1$ i.e.
$M_{\Phi_1}$ is same as the standard model up to zeroth order in
$M_W/M_{\Phi_1}$. Therefore, to this order, the vev of $\Phi_2$
(which is the equivalent of the standard model Higgs) will be
real, and there will be no spontaneous CP violation in the
theory (to order $M_W/M_{\Phi_1}$). This again proves that in the
limit of zero FCNC, there will be no SCPV. In appendix A, we
give explicit calculations in the mass basis that substantiates this 
conclusion.

 This result can be generalized to the case of arbitrary number
of Higgs doublets. For example for the case of three doublets, the
argument is that as long as all the doublets couple to quark
fields, at least two of the neutral Higgs bosons i.e. $H_{1,2}$
must be heavy in order to avoid large FCNC effects and this
implies that $|\mu^2_{23,}|\gg v^2_{wk}$; in that case their vev's
must be suppressed and  of order $\frac{v^3_{wk}}{|\mu^2_{2,3}|}$
and therefore small. The potential will then be forced to choose
the minimum such that all SCPV phases are zero.
                                                               
\section{High Scale Spontaneous CP violation leading to complex 
CKM while avoiding FCNC: Model with Extra Higgs Only}
In this section, we show how the FCNC
problem is avoided if spontaneous violation of CP symmetry arises
at a high scale. First we discuss this  using a model with two
$SU(2)_L\times U(1)_Y$ Higgs doublets $\phi_{1,2}$ as before and a
complex singlet $\sigma$. The potential for this case can be
written as follows:
\begin{eqnarray}
V_{\phi_1,\phi_2,\sigma}~=~V(\phi_{1,2})+~V(\sigma)+~V
(\phi,\sigma)
\end{eqnarray}
where $V(\phi_{1,2})$ is defined in Eqs.(\ref{eq1}), ( \ref{eq2}) and  
the other two terms are given by
\begin{eqnarray}
V(\sigma)~=~-M^2_0\sigma^*\sigma + M^2_1\sigma^2 +\lambda_\sigma
(\sigma^*\sigma)^2+\lambda'_\sigma \sigma^4 +\lambda''_\sigma
\sigma^3\sigma^* +h.c.
\end{eqnarray}
and
\begin{eqnarray}
V(\phi,\sigma)~=~M_{2,ab}\phi^\dagger_a\phi_b\sigma~+~\kappa_{1,ab}
\phi^\dagger_a\phi_b\sigma^2+ \kappa_{2,ab}
\phi^\dagger_a\phi_b\sigma^*\sigma+ h.c.\label{sigma}
\end{eqnarray}
It is clear that the minimum of the potential $V(\sigma)$ corresponds
to $<\sigma>=\Lambda e^{i\alpha}$, where $\Lambda\sim M_{0,1,2}\gg
v_{wk}$ and $\alpha$ can be large. Substituting this vev in the 
potential, we 
can write the
effective tree level potential for the $\phi_{1,2}$ fields at low
energies to be:
\begin{eqnarray}
V_{eff}(\phi_1,\phi_2)~=~V(\phi_{1,2})+V_{new}
\end{eqnarray}
where $V_{new}~=~(M_{2,ab}\Lambda
e^{i\alpha}+\kappa_{1,ab}\Lambda^2e^{2i\alpha}+
\kappa_{2,ab}\Lambda^2) \phi^\dagger_a\phi_b +h.c.\equiv
\Lambda^2(\lambda_{11} \phi^\dagger_1\phi_1+\lambda_{22}
\phi^\dagger_2\phi_2+\lambda_{12} e^{i\beta}
\phi^\dagger_1\phi_2) + h.c.$ If we keep only the neutral components
of the Higgs doublets, then the form of the potential is
\begin{eqnarray}
V_{eff}~=~\Lambda^2(\lambda_{11} \phi^\dagger_1\phi_1+\lambda_{22}
\phi^\dagger_2\phi_2+\lambda_{12} e^{i\beta}
\phi^\dagger_1\phi_2+h.c.)
+\sum \lambda_{abcd} \phi^\dagger_a\phi_b\phi_c\phi_d+h.c.
\end{eqnarray}
where $\Lambda \gg v_{wk}$. It is clear that although CP is spontaneously 
broken at a high scale $\Lambda$, at low energies, one has CP explicitly 
softly broken \cite{rebelo} by the bilinear terms in $\lambda_{12}$. Note 
that both the fields $\phi_{1,2}$
have Yukawa couplings and we can make a redefinition of the phase of
one of the doublet fields (say $\phi_2$) i.e. $\phi_2\rightarrow
e^{-i\beta}\phi_2$ so that all the bilinear and $O(\Lambda^2)$
terms in the potential become phase independent but the Yukawa
couplings become complex. Thus the effective theory at low
energies looks naively like hard CP violation, even though it is
spontaneous CP violation at a very high scale. The Yukawa
coupling Lagrangian looks like
\begin{eqnarray}
{\cal L}_Y~=~\bar{Q}_{La}(h^{u,1}_{ab}\phi_{1}~+~ 
h^{u,2}_{ab}e^{-i\beta}\phi_{2})u_{R,b}+
\bar{Q}_{La}(h^{d,1}_{ab}\tilde{\phi}_{1}~+
h^{d,2}_{ab}e^{i\beta}\tilde{\phi}_{2})~d_{R,b}+ h.c.
\end{eqnarray}
This still does not imply a viable complex CKM matrix; to
achieve that, we must show that the vev of $\phi_2$ where the
phase resides, does not become very tiny when we demand the 
suppression of FCNC. In order 
to show this, let us write down the extremization of the potential
as in section 2: For simplicity, we keep only the
$\lambda_{1111}$, $\lambda_{2222}$ and $\lambda_{1122}$ terms in
the potential but our results follow in general:
\begin{eqnarray}
(-\mu^2_1+\Lambda^2\lambda_{11}+\lambda_{1111}v^2_1+\lambda_{1122}
v^2_2)v_1+ v_2(\Lambda^2\lambda_{12}) =0\\
(-\mu^2_2+\Lambda^2\lambda_{22}+\lambda_{2222}v^2_2+\lambda_{1122}
v^2_1)v_2+ v_1(\Lambda^2\lambda_{12}) =0.
\end{eqnarray}
From these two equations, we find that both $v_{1}$ and $v_2$ are in 
general
of the same order regardless of what the neutral Higgs masses are.
This gives the CKM CP violation. As far as the masses of the
neutral Higgs fields go, we can fine tune one set of parameters to
keep one Higgs field light i.e. $\ll \Lambda$ and another will
remain heavy thus suppressing the FCNC effects.

Of course we do not need to make the rephasing $\phi_2\rightarrow 
e^{-i\beta}\phi_2$ and eliminate the phase from the bilinear terms.
If we do not do rephasing, the extremum equation of the Higgs potential 
would look like:
\begin{eqnarray}
-\Lambda^2\lambda_{12}v_1v_2 sin(\beta+\delta)-sin\delta \left[2\lambda_4 
v^2_1v^2_2 cos\delta+ v_1v_2(\lambda^2_5v^2_1+\lambda^2_6 v^2_2)\right]=0
\end{eqnarray}
Since $\Lambda^2\gg v^2$, it is clear that to an excellent approximation 
one has:
\begin{eqnarray}
\beta~=~-\delta.
\end{eqnarray}
The phase $\delta$ would then appear in the quark mass matrices which 
will be nontrivially complex, thus leading to a complex CKM matrix.
In Appendix B, we discuss how the fine tuning needed to keep the standard 
model Higgs at the electroweak scale does not prevent the components of the 
extra Higgs become superheavy in order to suppress the FCNC effects.

\section{SCPV without FCNC problem in fermionic extensions of
standard model} In this section, we briefly review the model in
\cite{branco} in which an extension of the standard model with a
$SU(2)_L$ singlet quark and a singlet Higgs field was presented
where one can have spontaneous violation of CP at high scale
without FCNC problem but with complex CKM. In this case, one
extends the standard model by the introduction of one singlet
vector like fermion of down type: $(D_{L,R})$ with $U(1)_Y$
quantum number $-2/3$ and a complex singlet Higgs field $\sigma$
as in sec. 3. The potential for the $\sigma$ field is the same as in
 equation (\ref{sigma}). As a result, the $\sigma$ field has a
complex vev leading to high scale spontaneous CP violation (since
$<\sigma>=\Lambda \gg v_{wk}$.

The CP violation is transmitted to the weak scale via its
couplings given below:
\begin{eqnarray}
{\cal L}_{\sigma}~=~\sum_a \bar{D}_Ld_{a,R}(g_a
\sigma+g'_a\sigma^*)+ ((f \sigma+f'\sigma^*)\bar{D}_LD_R+h.c.
\end{eqnarray}
where $g_{a},g'_a,f,f'$ are real due CP conservation. But after
symmetry breaking, the mass matrix contains terms mixing the heavy $D$ 
quarks
with the light $d$ quarks \cite{branco}. This can be seen by
writing down the full down quark mass matrix (in the notation
$\bar{\psi}_LM_{dD}\psi_R$):
\begin{eqnarray}
M_{dD}~=~\pmatrix{m_d & 0\cr \Lambda(ge^{i\delta}+g'e^{-i\delta})
& \Lambda(fe^{i\delta}+f'e^{-i\delta})}
\end{eqnarray}
where $g$ and $g'$ denote the row vectors $(g_1,g_2,g_3)$ and
$(g'_1,g'_2,g'_3)$. Diagonalizing $M_{dD}M^\dagger_{dD}$, we can
get the generalized $4\times 4$ CKM matrix which indeed has a
complex phase in the $3\times 3$ sector involving the standard model 
quarks even in the limit of heavy $D$ quark masses. This is an example of
a breakdown of the decoupling theorem \cite{branco}. Clearly since
there is only one neutral Higgs boson coupling to the effective
down quark mass matrix, there is no FCNC effects at the tree level
as in the case of the standard model. Clearly, if the masses of
the vectorlike quarks were at the weak scale, the mixing between
the light $d$ quarks and $D$ would be significant and lead to
large FCNC effects at low energies.

 This provides a second way
to introduce spontaneous CP violation without simultaneously
having flavor changing neutral current effects. Note that the common
thread between the examples in sec. 3 and 4 is the fact that CP is 
violated
spontaneously at high scale, which highlights the main point of
this paper. In the remainder of this paper, we show how these
ideas can be embedded into extended models on the way towards a
possible grand unified scheme where spontaneous CP violation
occurs at the GUT scale.

\section{Embedding high scale SCPV into left-right symmetric models}
The left-right symmetric models are based on the gauge group
$SU(2)_L\times SU(2)_R\times U(1)_{B-L}\times SU(3)_c$ with
fermions assigned in a left-right symmetric manner\cite{lrs} and
Higgs belonging to bidoublet field $\Phi(2,2,0)$ and a pair of
fields of either $(\chi_L(2,1,-1)\oplus \chi_R(1,2,-1))$ type
(called $\chi$-type below) or
$(\Delta_L(3,1,+2)\oplus\Delta_R(1,3,+2))$ type (called
$\Delta$-type below).
 The left-right symmetric models are ideally suited to embed the first class of
high scale SCPV models since the bidoublet Higgs field already
contains the necessary two standard model doublet Higgs fields in
it. All we have to do is to embed the high scale singlet field
into a left-right Higgs field. We present two different ways to do this 
embedding in the two subsections below:

\subsection{Left-Right SCPV: Model I}
The first way to implement high scale SCPV is by choosing two pairs of 
$\chi$ type or $\Delta$-type fields. Two
pairs are needed since with a single pair, constraint that $W_R$
scale must be much higher than $W_L$ scale suppresses the SCPV
phase by a factor $M_{W_L}/M_{W_R}$\cite{masiero}. The two
$\Delta$ type model has been discussed in \cite{basecq} where 
at the high scale, the $\Delta_R$'s have vevs as follows:
$<\Delta^0_{1,R}>~=~v_{1,R}$ and
$<\Delta_{2,R}>~=v_{2,R}e^{i\delta}$. The coupling of the form
$Tr(\Phi^\dagger\tau_2\Phi^*\tau_2) Tr(\Delta^\dagger_{1,R}\Delta_{2,R})$
then induces the term
$\lambda_{12}e^{i\delta}\Lambda^2\phi^*_1\phi_2$ at low energies
and the rest of the discussion is as in section 3 above.

Let us now turn our attention to embedding of the model of Ref.\cite{branco} 
into the left-right model. We consider the
left-right model without the bidoublet but with the
$(\chi_L(2,1,-1)\oplus \chi_R(1,2,-1))$ pair and three pairs of
$SU(2)_L\times SU(2)_R$ singlet vector-like quarks $(P_{L,R}(1, 1,
4/3)$ and $N_{L,R}(1,1,-2/3)$). Such models were extensively
studied in the early 90's but not from the point of view of
spontaneous CP violation\cite{babu}. We take a complex singlet
Higgs field $\sigma$ as before and assume the theory to be CP
conserving prior to symmetry breaking so that all couplings in the
theory are real. Again, we assume the potential for the $\sigma$
field to be as in Eq.\ref{sigma} so that its minimum corresponds
to a complex vev for $<\sigma>= \Lambda e^{i\delta}$ as before.
The vevs for the fields $\chi_{L,R}$ are real.

To study the implications of the theory for low energy quark
mixings, let us write down the quark Yukawa couplings:
\begin{eqnarray}
{\cal L}_Y~=~h^u_{ab}[\bar{Q}_{L,a}\chi_LP_{R,b}+
\bar{Q}_{R,a}\chi_RP_{L,b}]+
h^d_{ab}[\bar{Q}_{L,a}\tilde{\chi}_LN_{R,b}+
\bar{Q}_{R,a}\tilde{\chi}_RN_{L,b}] + h.c.\\ \nonumber
+[f^u_{ab}\sigma +f^{u,'}\sigma^*]\bar{P}_{L,a}P_{R,b}+
[f^d_{ab}\sigma +f^{d,'}\sigma^*]\bar{N}_{L,a}N_{R,b}] + h.c.
\end{eqnarray}
After spontaneous symmetry breaking we have $<\sigma>=\Lambda
e^{i\delta}$, $<\chi^0_{L,R}>= v_{L,R}$ with $v_R\sim \Lambda \gg
v_L$. This leads to the mass matrix of the form:
\begin{eqnarray}
{\cal M}_{uP}~=~\pmatrix{0 & h^u_{ab}v_L\cr h^u_{ba}v_R & M_P}\\
{\cal M}_{dN}~=~\pmatrix{0 & h^d_{ab}v_L\cr h^d_{ba}v_R & M_N}
\end{eqnarray}
Left-right symmetry requires that $M_{P,N}=M^\dagger_{P,N}$
whereas the matrices $h^{u,d}$ are real. After diagonalization,
the effective up and down quark mass matrices become:
\begin{eqnarray}
M_{u,d}\simeq v_Lv_Rh^{u,d,T}M^{-1}_{P,N}h_{u,d}
\end{eqnarray}
These matrices are hermitean and therefore lead to equal left and
right handed CKM matrices as in the usual left-right models with
bi-doublets and lead to complex CKM matrices. In fact one can
write the rotation matrices for both the up and down sector as
follows in a basis where the couplings $h^{u,d}$ are diagonal:
\begin{eqnarray}
V^{u,d}~=~M^{-1/2}_{u,d}h^{u,d}U_{P,N}{M^{diag}}^{-1/2}_{P,N}\sqrt{v_Lv_R}
\end{eqnarray}
Clearly since $U_{P,N}$ is a unitary matrix with complex phases,
$V^{u,d}$ will lead to complex CKM matrix i.e.
$U_{CKM}=V^uV^{d,\dagger}$.

As far as the FCNC effects are concerned, they arise only in order
$m_{u,d}/M_{P,N}$ and therefore suppressed when $\Lambda\to $
large values. Note however that the quark mixing effects arise in
zeroth order of this parameter.

\subsection{Left-Right SCPV Model II:Connecting the CP violation 
and seesaw scales}
In this subsection, we present a more economical left-right embedding of the 
high scale spontaneous CP violation with suppressed  FCNC. The model 
consists of the usual left-right assignment of the fermions\cite{lrs} and 
Higgs system consists of a single bidoublet $\phi(2,2,0)$ and the 
$\chi_L(2,1,-1)\oplus\chi_R(1,2,-1)$. Here spontaneous CP violation is 
implemented via the vev of a CP and P odd real singlet scalar field 
$\eta$\cite{cmp}.
The CP invariant Higgs potential for the theory can be written as:
\begin{eqnarray}
V(\chi_{L,R},\eta,\phi)~=~V_0(\phi)+ i\mu 
\eta Tr(\phi^\dagger_1{\phi}_2)+ M'\chi^\dagger_L\phi\chi_R+V_2(\eta, 
\chi_{L,R})
\end{eqnarray}
where
\begin{eqnarray}
 V_0(\phi)~=~-\mu^2_{ab}Tr(\phi^\dagger_a\phi_b)+
\sum\kappa_{abcd}Tr(\phi^\dagger_a\phi_b\phi^\dagger_c\phi_d)+
\kappa'_{abcd}Tr(\phi^\dagger_a\phi_b)Tr(\phi^\dagger_c\phi_d)+h.c.
\end{eqnarray}
with (a,b) going over (1,2) with $\phi_1=\phi$ and 
$\phi_2=\tau_2\phi^*\tau_2$.
\begin{eqnarray}
 V_2(\eta, \chi_{L,R})~=~M^2_\eta \eta^2+\lambda_\eta \eta^4
-M^2_\chi (\chi^\dagger_L\chi_L+ \chi^\dagger_R\chi_R)+
\lambda_\chi (\chi^\dagger_L\chi_L+ \chi^\dagger_R\chi_R)^2+\\ \nonumber
\lambda'_\chi(\chi^\dagger_L\chi_L- \chi^\dagger_R\chi_R)^2+
M'_\eta\eta (\chi^\dagger_L\chi_L- \chi^\dagger_R\chi_R)
\end{eqnarray}
We have assumed that under CP transformation $\eta\rightarrow -\eta$ and 
$\chi_L\rightarrow \chi^\dagger_R$ and $\phi\rightarrow \phi^\dagger$. 
Invariance under this transformation requires that all parameters in the 
potential be real (except for one imaginary coupling shown explicitly in 
the above equation).

Note now that if the term in the potential connecting $\eta$ and $\chi$ 
fields was absent, we would have $<\eta>=0$ since $M^2_\eta >0$. However 
as soon as $SU(2)_R$ symmetry is broken by $<\chi^0_R>\neq 0$, the 
$M'_\eta$ term in the potential introduces a tadpole term for $\eta$ 
thereby generating 
\begin{eqnarray}
<\eta>~\simeq \frac{+M'_\eta v^2_R}{2M^2_\eta}.
\end{eqnarray}
Since $\eta$ is CP odd, 
this breaks CP spontaneously. The way it 
manifests is that the $i\mu<\eta> Tr(\phi^\dagger_1\phi_2)$ term now 
combines with the $\mu^2_{12}Tr\phi^\dagger_1\phi_2$ to generate at low 
energies an effective soft CP breaking term as in Eq. (13)
 where $\phi_{1,2}$ are the two doublets contained in the bidoublet 
$\phi$ of the left-right model. The same
arguments as in the Appendix B 
then guarantee that in this model the FCNC can be suppressed by making 
one of the left-right Higgs doublets superheavy.

This can also be seen in an alternative manner by minimizing 
the potential, noting that there is a range of values of 
the parameters in the potential for which we have
$<\chi^0_R>~=v_R\neq 0; <\eta>\neq 0; <\chi_L>=0$ provided $M'_\eta 
<\eta> > 2\lambda'_\chi u^2$. The vevs 
of $\chi_R$ and $\eta$ fields are much larger than the weak scale.

 An interesting point worth stressing is that in this model, the scale of CP 
violation and the seesaw\cite{seesaw1} scale for neutrino masses are 
connected. To see 
this, note that the right handed neutrino masses come from the higher 
dimensional term $(L_R\bar{\chi}_R)^2/M_{Pl}$ leading to seesaw right 
handed neutrino masses given by $M_{seesaw}\simeq\frac{v^2_R}{M_{Pl}}$ 
and from Eq.(29), the CP violating scale $<\eta>$ and $M_{seesaw}$ owe 
their origin to the same scale $v_R$ i.e. violation of parity. Since in 
grand unified theories, $v_R$ can be identified with the GUT scale, one 
would therefore relate several scales of the theory i.e. $M_{SCPV}$, 
$M_{seesaw}$ and $M_{GUT}$.

\section{Possible extensions to supersymmetry and SUSY CP problem}
As is well known, generic minimal supersymmetric extensions of the 
standard model (MSSM) are plagued with the SUSY CP problem. There have 
been many solutions suggested to solve this problem\cite{cures}.
A simple solution to this problem would of course be to have CP 
spontaneously broken. However, in MSSM, CP cannot be spontaneously 
broken. Furthermore, it has also been pointed out that \cite{masip} it is 
particularly hard to have 
spontaneous CP breaking by considering multi-Higgs generalizations of the 
MSSM. A possibility for achieving spontaneous CP breaking within SUSY involves 
the introduction of singlet chiral fields\cite{teixeira}. As far as the 
FCNC 
effects are concerned, in these models one may 
fine tune the $\mu$ terms to make of the 
extra Higgs doublets heavy thereby eliminating large FCNC effects.
However, the early versions of these models are no longer viable since 
they had a real CKM matrix, in contradiction with recent experimental data.

Therefore, the ideas described in this paper may be particularly
useful if one wants to solve the SUSY CP problem by spontaneous CP 
violation in a viable scenario, where vacuum phases do lead to
a complex CKM matrix, while at the same time suppressing FCNC effects. 
In fact, recently it has been suggested one such model which includes 
two singlet Higgs superfields and adds an extra vector like singlet fermion 
to MSSM\cite{romao} to break CP spontaneously and generate a complex CKM matrix. 
One can embed this scheme into the SUSY left-right model. Detailed analysis of SUSY 
models that exploit the ideas of this paper is under way and will be 
taken up in a forthcoming publication.

\section{Conclusion}
We have emphasized the close connection between 
spontaneous CP violation and FCNC effects in theories 
where CP breaking vev is at the weak scale. We have also shown that in 
order to avoid FCNC effects while at the same time generating a complex 
CKM matrix through vecuum phases, one is naturally led to have 
spontaneous CP breaking at a high energy scale, well above the 
electroweak scale.We then describe two classes 
of models one without and one with extra heavy fermions where having a 
high vev break CP spontaneously leads to complex 
CKM matrix as given by experiment without simultaneously having large 
FCNC effects. We then show how these models can be embedded into the high 
scale left-right models where parity violation and neutrino mass are 
connected via the seesaw mechanism. We find one particular model where 
spontaneous parity violation triggers the spontaneous CP violation thus 
connecting the three scales: seesaw scale for neutrino masses, scale of 
spontaneous parity and CP violation. 

In conclusion, if our view on the origin of CP violation is correct, then 
small neutrino masses and CP violation at low energies would have in 
common the fact that they are both manifestations of physics occuring at 
very high energy scale.

 \newpage

\begin{center}
{\bf Appendix A}
\end{center}

In this appendix, we elaborate on the connection between SCPV and FCNC 
and complex CKM in a two Higgs doublet model. For this purpose, we write 
the Yukawa Lagrangian as:
\begin{eqnarray}
{\cal L}_{Y}~=~\sum_{a,b} (h^{u,i}_{ab}\bar{Q}^0_{La}\phi_iu^0_{R,b}+ 
u\rightarrow d)~+~h.c.
\end{eqnarray}
It can be readily seen \cite{lavoura}, \cite{gcb} that in the quark mass 
eigenstate 
basis, the scalar coupling can be written as:
\begin{eqnarray}
{\cal L}_{scalar}~=~\left[\bar{u}D_uu+\bar{d}D_dd\right]\frac{H}{v}
-\left[\bar{u}(N_uP_R+N^\dagger_uP_L)u+\bar{d}(N_dP_R+N^\dagger_dP_L)d
\right]   
\frac{R}{v}\\ \nonumber 
+i\left[\bar{u}(N_uP_R-N^\dagger_uP_L)u-\bar{d}(N_dP_R-N^\dagger_dP_L)d
\right]   \frac{I}{v}
\end{eqnarray} 
where 
\begin{eqnarray}
H~=~\frac{1}{v}\left[v_1R_1+v_2R_2\right]\\ \nonumber 
R~=~\frac{1}{v}\left[v_2R_1-v_1R_2\right]\\ \nonumber
I~=~\frac{1}{v}\left[v_2I_1-v_1I_2\right] 
\end{eqnarray}
with $\phi^0_1~=~\frac{1}{\sqrt{2}}\left[v_1+R_1+iI_1\right]$
$\phi^0_2~=~\frac{1}{\sqrt{2}}e^{i\delta}\left[v_2+R_2+iI_2\right]$.
where
\begin{eqnarray}
N_d~=~U^\dagger_{dL}\left[\frac{v_2}{\sqrt{2}}Y^d_1-\frac{v_1}{\sqrt{2}} 
e^{i\delta}Y^d_2\right]U_{dR}
\end{eqnarray}
where $U_{d_{L,R}}$ are the unitary matrices which diagonalize the down 
quark mass matrix $M_d$. Analogous expressions are there for $N_u$. It is 
clear that $N_{d,u}$ are in general not diagonal and therefore $R$ and 
$I$ mediate FCNC.

The quark mass matrices are in the form   
\begin{eqnarray}
M_dM^\dagger_d~=~H_{real}+ 
2iv_1v_2sin\delta(Y^{d2}{Y^{d1}}^T-Y^{d1}{Y^{d2}}^T)
\end{eqnarray}
where $H_{real}$ is a symmetric real matrix. It is clear 
that $M_dM^\dagger_d$ (and similarly $ M_uM^\dagger_u$) is an arbitray 
complex matrix and therefore CKM is a complex matrix. 

If one fine tunes such that $Y^d_1 \propto Y^2_d$, $N_d$ would be 
diagonal and FCNC would be eliminated. But in that case, Eq(34) implies 
that $M_dM^\dagger_d$ becomes real. This illustrates the connection 
between FCNC and the possibility of generating a complex CKM by a 
vacuum phase.

\begin{center}
{\bf Appendix B}
\end{center}
In this appendix, we discuss how the extra neutral Higgs fields in the 
model of section III that are potential mediators of FCNC effects can be 
made heavy while at the same time the SM Higgs can be kept light by one 
fine tuning. We will work with the potential in Eq.(13,14).Clearly, the 
minimum of this potential corresponds to:
\begin{eqnarray}
<\phi_1>~=~\pmatrix{0\cr v_1}; <\phi_2>~=~\pmatrix{0\cr v_2e^{i\delta}}
\end{eqnarray}
 Let us work in a basis in which
\begin{eqnarray}
\pmatrix{H_1 \cr H_2}~=~\frac{1}{v}\pmatrix{v_1 & v_2\cr v_2 & 
-v_1}\pmatrix{\phi_1 \cr e^{-i\beta}\phi_2}
\end{eqnarray}
The potential in Eq. (13) looks as follows:
\begin{eqnarray}
V(H_{1,2})~=~\Lambda^2\left(\lambda_{11}H^\dagger_1H_1+ 
\lambda_{22}H^\dagger_2H_2+(\lambda_{12}H^\dagger_1H_2+ h.c.) 
\right)
+\lambda_1 (H^\dagger_1 H_1)^2 +\lambda_2 (H^\dagger_2 H_2)^2\\ 
\nonumber
+\lambda_3 (H^\dagger_1 H_1)(H^\dagger_2H_2)
+\lambda_4 (H^\dagger_2 H_1)(H^\dagger_1H_2)+\left[\lambda_5 
H^\dagger_1H_2+\lambda_6 
H^\dagger_1H_1+\lambda_7H^\dagger_2H_2\right]H^\dagger_1H_2 + h.c.
\end{eqnarray}
Even though we use the same $\lambda$'s in both Eq.(13) and here, they 
are different and in fact now $\lambda_{12},\lambda_{5,6,7}$ are in 
general complex while the other $\lambda$'s are real.

Now we can write the $H_{1,2}$ in terms of their components:
\begin{eqnarray}
H_1~=~\pmatrix{G^+\cr \frac{1}{\sqrt{2}}(v+H+iG)}; H_2~=~\pmatrix{C^+\cr 
\frac{S+iP}{\sqrt{2}}}
\end{eqnarray}

As already discussed in \cite{book}, the stability of vacuum demands 
that, the coefficients of the linear terms in $(H, S, P)$ vanish and 
gives 
\begin{eqnarray}
\Lambda^2 \lambda_{11}+2\lambda_1 v^2 =~0\\ \nonumber
\Lambda^2 \lambda_{12}+\lambda_6 v^2 =~0
\end{eqnarray}
These are the fine tuning conditions in the $(H_{1,2})$ basis to have SM 
Higgs field light and have the correcxt electroweak symmetry breaking.
We can now write down the mass matrix for the other neutral Higgs fields 
$(H, S, P)$ as follows\cite{book}:
\begin{eqnarray}
{\cal M}_{H,S,P}~=~\pmatrix{4v^2\lambda_1 & 2v^2Re\lambda_6 & -2v^2Im 
\lambda_6\cr 2v^2Re\lambda_6 & \lambda_2 \Lambda^2+(\lambda_3 
+\lambda_4+2 Re 
\lambda_5)v^2 & -2v^2Im\lambda_5\cr  -2v^2Im \lambda_6 & -2v^2Im\lambda_5 
& \lambda_2\Lambda^2 +\lambda_3 v^2+(\lambda_4-2Re\lambda_5)v^2}
\end{eqnarray}
From this expression, we can explicitly see that the beyond the 
standard model neutral Higgs particles $(S,P)$ have masses of order 
$\Lambda$ whereas the SM Higgs field has mass of order of the elctroweak 
scale. Also the mixings of the SM Higgs which can generate FCNC effects 
are of order $v^2/\Lambda^2$ and hence very small as $(S,P)$ are made 
heavy. Also $\lambda_2\Lambda^2 +\lambda_3 v^2$ gives the mass of the 
charged Higgs field from the second Higgs field $H_2$. Thus we have 
complex CKM from SCPV while at the same time suppressing the FCNC  
effects.

 The work of  R. N. M. is supported by the National
Science Foundation grant no. Phy-0354401 and the work of GCB is  
supported by Fundacao 
para a Ciencia e a Tecnologia (FCT, Portugal), through the projects 
POCTI/FNU/44409/2002, PDCT/FP/FNU/50250/2003, POCI/FP/63415/2005, 
POCTI/FP/FNU/50167/2003, which are partially funded through POCTI (FEDER).
Both the authors are 
very grateful for the Alexander von Humboldt Senior Research Award which made 
this collaboration possible. They are also grateful to A. Buras 
and M. Lindner at TUM and R. N. M. to H. Fritzsch at 
LMU for kind hospitality when the work was done.

\end{document}